\documentclass[preprint,showpacs]{revtex4}%
\usepackage[T2A]{fontenc}
\usepackage[cp1251]{inputenc}
\usepackage[english]{babel}
\usepackage{amsmath}
\usepackage{amsfonts}
\usepackage{graphicx,epsfig}
\usepackage{subfigure}
\usepackage{amssymb}
\usepackage[indent,small]{caption2}
\usepackage{graphicx}%
\providecommand{\U}[1]{\protect\rule{.1in}{.1in}}
\begin{document}
\title{Bloch Oscillations in the Optical Waveguide Array.}
\author{Gozman M. I.$^{1,2},$Polishchuk I. Ya.$^{1,2}$.}
\affiliation{$^{1}$ RRC Kurchatov Institute, Kurchatov Sq., 1, 123182 Moscow, Russia}
\affiliation{$^{2}$ Moscow Institute of Physics and Technology, 141700, 9, Institutskii
per., Dolgoprudny, Moscow Region, Russia}

\begin{abstract}
The multiple scattering formalism is proposed describing the
guided modes in the optical waveguide array within the framework
of macroscopic electrodynamics. It is shown that, under
sufficiently general assumptions, our approach justifies the
phenomenological model used widely to treat various physical
phenomena in the optical micro- and nano-structures. It is found
that the theory developed in this paper describes the real
experiments in which the the Bloch oscillations are observed.
Surprisingly, not only qualitative but also reasonably
quantitative agreement is found.

\end{abstract}

\maketitle


\section{Introduction.\label{Sec:Intro}}

Various artificial materials such as metamaterials \cite{meta}, photonic
crystals \cite{Joannopoulos, Busch}, and waveguide arrays \cite{Longhi} are
considered as promising structures to manipulate light effectively. It is
their spatial periodicity that connects these artificial materials and
conventional crystals (metal, semiconductors, dielectrics). For this reason,
the Bloch functions inherent in the Schr\"{o}dinger equation with a periodic
potential should be typical for the solutions of the Maxwell equations with
the periodical dependence of the refractive index. Hence, many physical
effects inherent in solid state physics should have their optical counterparts
\cite{Longhi}.

For the Schr\"{o}dinger equation, the electron Bloch functions describe the
propagating state. It is known that a superimposition of a disorder or an
external electrical field may result in the spatial localization of an
electron in solids. In the first case, the electron experiences either
Anderson localization or dynamic localization. In the second case, an electron
experiences the Bloch oscillations and Zener tunneling. In this paper, we
investigate the optical analog of the electron Bloch oscillations.

In the standard the first principle solid state physics the description of the
phenomena is based on the analysis of the corresponding Schr\"{o}dinger
equation. Within the quasi-classical description, the Bloch oscillations are
known as a finite- oscillation motion of an electron in the periodical
potential when the external dc electric field is applied. However, such finite
motion can be understood setting on the exact solution of the Schr\"{o}dinger
equation as follows. A particle moving in a periodical potential possesses the
infinite Bloch state which belongs to the continuous energy spectrum. The
application of the external dc field changes the energy spectrum drastically.
The electron states become spatially finite, belonging to the discrete energy
spectrum called a Wannier-Stark ladder. The wave function for each state of
the ladder manifests a localized state, while a wave packet of these functions
describes the Bloch oscillations. In general, the period of these oscillations
is much larger than either the scattering time of electron with impurities or
the Zener tunneling time. For this reason, the Bloch oscillations are never
observed in the conventional crystal. In fact, the first observation of the
Wannier-Stark ladder \cite{StarcLedder} and the Bloch oscillations
\cite{Superlatticie} has become possible using the semiconductor superlattices
in which the shorter Bloch oscillations period was attained. Also the Bloch
oscillations have experimentally been observed for cold atoms and the
Bose-Einstein condensates in optical lattices \cite{coldatom}.

The optical waveguide array (OWA)\ considered in this paper enables us a
visualization of the Bloch oscillations in the spatial domain as an
oscillatory light beam path. Each waveguide is a homogeneous one and serves as
an attractive atomic potential in the crystals. The dc field is mimicked by
the monotonic change of the refractive index of the waveguides as one passes
from one waveguide to another. The effect is reached, in particular, by
applying the temperature gradient across the thermooptical material \cite{87,
92, Pertsch2}, by a suitable change of the waveguide geometrics \cite{86}, or
by a circularly-curving the waveguides \cite{88, 91, 93}.

\begin{figure}[ptbh]
\centering
\includegraphics[width=0.8\textwidth]{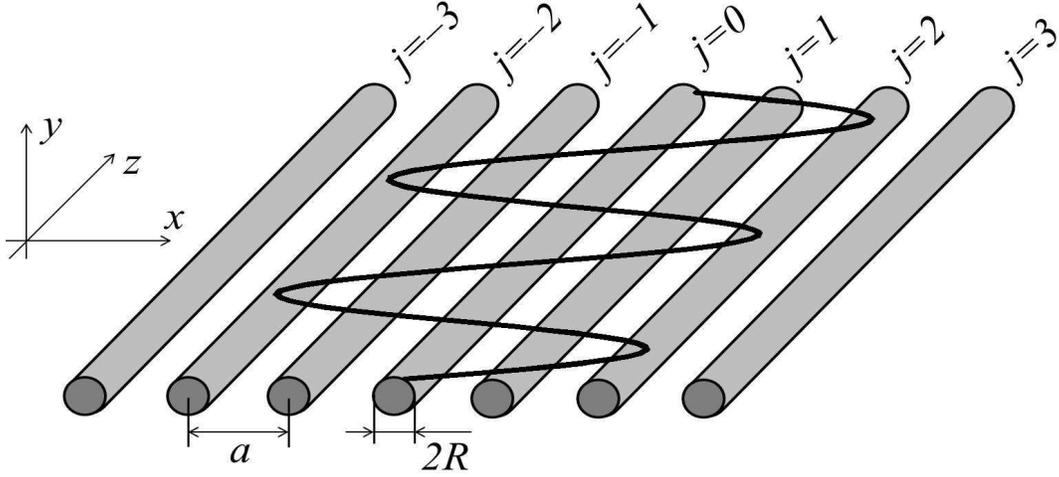}\caption{The OWA. The solid
line manifests the path of the light beam inherent in the Bloch oscillations.}%
\label{fig1}%
\end{figure}

To describe the optical Bloch oscillations in the array of parallel waveguides
(see. Fig. \ref{fig1}), a very viable phenomenological model was proposed in
\cite{PertschTheor}. Along with the optical Bloch oscillations, this model was
used to investigate some other various physical effects in the optical
structures such as nonlinear Bloch oscillations, the Bloch oscillations in the
waveguide arrays with the second-order coupling, the Bloch-Zener oscillations
in optical waveguide ladders and binary superlattices, the gradon localization
\cite{Longhi, Pertsch2, Wang1, Wang2, Dreisow1, Dreisow2, PRA2010}. Within
this model, a set of modal amplitudes $a_{j}\left(  z\right)  $ are introduced
which describe a behavior of the effective light amplitude along the $j$-th
waveguide. According to Ref. \cite{PertschTheor}, these amplitudes obey the
system of coupled equations
\begin{equation}
\left(  i\frac{\partial}{\partial z}+\alpha\cdot j\right)  a_{j}+\gamma\left(
a_{j+1}+a_{j-1}\right)  =0.\label{intro}%
\end{equation}
Here $z$ is the direction along the waveguide axis, the index $j=0,\pm1,...$
determines the position of the waveguide in the array. The first term in Eq.
(\ref{intro}) describes the propagation of light beam along the isolated
waveguide, the parameter $\alpha$ being responsible for the refractive index
ramp. The second term describes the influence of the nearest neighbor
waveguides on the light propagation, the parameter $\gamma$ being responsible
for this influence. The phenomenological constants $\alpha$ and $\gamma$
entering Eq. (\ref{intro}) remain unknown and are obtained only as a result of
comparison with an experiment.

To our knowledge, equations like (\ref{intro}) have never been derived within
the macroscopic electrodynamics approach. This approach assumes that the array
of the infinite homogeneous waveguides is considered and the refractive index
of each waveguide is known.

A solution to the corresponding Maxwell equations describes a
distribution of the electromagnetic field in the whole space. For
the array to possess the guided properties, the field should
mainly be concentrated inside the array. Thus, we are interested
only in those solutions of the Maxwell equations for the waveguide
array which vanish as $\left\vert y\right\vert \rightarrow\infty$
(see. Fig. \ref{fig1}). These are evanescent modes. It stands to
reason, that the solution should possess a finite amplitude inside
the array. The goal of the paper is to determine correctly the
amplitudes $a_{j}\left(  z\right)  $, connecting them with a
superposition of these guided mode solutions of the Maxwell
equations. A direct derivation of Eq. (\ref{intro}) from the
Maxwell equations allows us to obtain the constants $\alpha$ and
$\gamma$ in terms of the workpiece geometrics and refractive
indices of the waveguides and to establish the validity range of
the model described by Eq. (\ref{intro}).

Like in quantum mechanics, for the waveguide array under consideration two
formulations of the problem are possible. The first one is the problem of
scattering of electromagnetic waves by the array. Such problem is investigated
in Refs. \cite{Felbacq1994, ZhangLi1998, Felbacq2000, DuLin2009}. The approach
used in these papers is based on the exact solution for the scattering of the
electromagnetic wave by a single infinite cylinder \cite{VanDeHulst}. The
solution for the scattering mode does not vanish as it escapes from the array.
The second task is similar to that of the bound states in quantum mechanics.
It is related with the guided modes inherent in the array. For a single
waveguide, the solution is found in \cite{Marcuse}. Finding the solution for
the waveguide array is just the goal of the paper. The approach we propose is
similar to the multisphere Mie scattering formalism developed in our previous
papers \cite{We2007, We2008, We2009, We2010} to describe the high-quality
guided modes in the arrays of spherical particles.

The paper is organized as follows. First, we derive the system of equations,
which describes the guided modes in the array of parallel dielectric
waveguides. For this purpose, we expand the electromagnetic field in the
vector cylinder harmonics. The system of equations obtained is a formally
exact one for the guided mode propagating in the array and consists of the
infinite number of equations. To make the problem solvable, we truncate it
using the nearest neighbor approximation and zero-harmonic approach. Under
these simplifications, one obtains that the amplitudes $a_{j}\left(  z\right)
$ obey Eq. (\ref{intro}). Finally, we apply the results obtained to treat the
real experiments \cite{87, Pertsch2}. Surprisingly, not only qualitative but
also reasonable quantitative agreement is found.

\section{Derivation of the equation for the guided modes}

\label{Sec:MCMSF}

Let us consider the array of $N$ parallel cylindrical dielectric waveguides
(see Fig. \ref{fig1}). The axes of the waveguides are in the $xz$-plane and
are parallel to the $z$-axis. The array is equidistant, $a$ being the distance
between the axes of the nearest waveguides. In this paper, the array of
infinite parallel cylinder waveguides is considered. All the waveguides are
assumed to possess the same radius $R_{j}=R$ but different refractive indices
$n_{j}$. It is assumed that the contrast between the nearest waveguides is
$n_{j}-n_{j-1}=\mathrm{const}$.

Suppose that a guided mode with a frequency $\omega$ is excited within the
array. Because of the translation invariance in the $z$ direction, all the
components of the electromagnetic field describing the guided mode depend on
the coordinate $z$ as $e^{i\beta z}$, $\beta$ being a propagation constant
(the wave vector component in the $z$-direction) of the guided mode. Thus, all
the components of the guided mode are proportional to the factor $e^{-i\omega
t}e^{i\beta z}$. Let us consider the guided mode inside and outside of the array.

\begin{figure}[ptbh]
\centering
\includegraphics[width=0.8\textwidth]{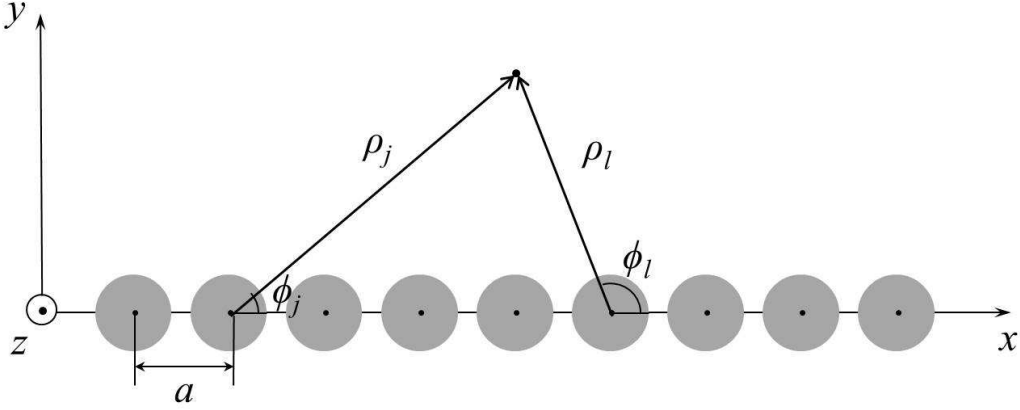}
\caption{The OWA. The polar coordinates of radius-vector $\mathbf{r}$ relative
to different waveguides.}%
\label{fig2}%
\end{figure}

Since the guided mode possesses a finite value, the electromagnetic field
inside of the $j$-th waveguide may be represented in the form%

\begin{equation}%
\begin{array}
[c]{l}%
\displaystyle\mathbf{\tilde{E}}_{j}(\mathbf{r})=e^{-i\omega t}\,e^{i\beta
z}\,\sum\limits_{m=0,\pm1...}e^{im\phi_{j}}\,\Bigl(c_{jm}\,\tilde{\mathbf{N}%
}_{\omega_{j}\beta m}(\rho_{j})-d_{jm}\,\tilde{\mathbf{M}}_{\omega_{j}\beta
m}(\rho_{j})\Bigr),\medskip\\
\displaystyle\mathbf{\tilde{H}}_{j}(\mathbf{r})=e^{-i\omega t}\,e^{i\beta
z}\,n_{j}\sum\limits_{m=0,\pm1...}e^{im\phi_{j}}\,\Bigl(c_{jm}\,\tilde
{\mathbf{M}}_{\omega_{j}\beta m}(\rho_{j})+d_{jm}\,\tilde{\mathbf{N}}%
_{\omega_{j}\beta m}(\rho_{j})\Bigr),\quad\rho_{j}<R.
\end{array}
\label{AppS_EHint}%
\end{equation}
Here $\mathbf{r}=\left(  x,y,z\right)  =\left(
\boldsymbol{\rho},z\right) ,\rho_{j}=\left\vert
\boldsymbol{\rho}-\mathbf{a}j\right\vert $, $\phi_{j}$ is the
polar angle of the vector $\boldsymbol{\rho}-\mathbf{a}j$ (see
Fig. \ref{fig2}), $\omega_{j}=n_{j}\omega$. The vector cylinder
harmonics
$\tilde{\mathbf{M}}_{\omega_{j}\beta m}(\rho_{j})$ and $\tilde{\mathbf{N}%
}_{\omega_{j}\beta m}(\rho_{j})$ are defined as follows%

\begin{equation}
\tilde{\mathbf{N}}_{\omega_{j}\beta m}(\rho_{j})=\mathbf{e}_{r}\,\frac{i\beta
}{\varkappa_{j}}\,J_{m}^{\prime}(\varkappa_{j}\rho_{j})-\mathbf{e}_{\phi
}\,\frac{m\beta}{\varkappa_{j}^{2}\rho_{j}}\,J_{m}(\varkappa_{j}\rho
_{j})+\mathbf{e}_{z}\,J_{m}(\varkappa_{j}\rho_{j}), \label{tildeN}%
\end{equation}

\begin{equation}
\tilde{\mathbf{M}}_{\omega_{j}\beta m}(\rho_{j})=\mathbf{e}_{r}\,\frac
{m\omega_{j}}{\varkappa_{j}^{2}\rho_{j}}\,J_{m}(\varkappa_{j}\rho
_{j})+\mathbf{e}_{\phi}\,\frac{i\omega_{j}}{\varkappa_{j}}\,J_{m}^{\prime
}(\varkappa_{j}\rho_{j}),\label{tildeM}%
\end{equation}
where $\varkappa_{j}=\sqrt{\omega_{j}^{2}-\beta^{2}},$ $J_{m}(\varkappa
_{j}\rho_{j})$ is the Bessel function, and the prime means the derivative with
respect to the argument $\varkappa_{j}\rho_{j}$. The functions $\tilde
{\mathbf{N}}$ and $\tilde{\mathbf{M}}$ are orthogonal. Thus, the guided mode
inside the $j$-th rod which possesses the frequency $\omega,$ is determined by
the propagation constant $\beta$ and by the set of the partial amplitudes
$c_{jm}$, $d_{jm}$.

Let us turn to the electromagnetic field for the same guided mode outside of
the array. Each waveguide of the array contributes to this field. The
contribution induced by the $j$-th waveguide and vanishing at $y\rightarrow
\pm\infty$ may be represented in the form%

\begin{equation}%
\begin{array}
[c]{l}%
\displaystyle\mathbf{E}_{j}(\mathbf{r})=e^{-i\omega t}\,e^{i\beta z}%
\sum\limits_{m}e^{im\phi_{j}}\,\Bigl(a_{jm}\,\mathbf{N}_{\omega^{\prime}\beta
m}(\rho_{j})-b_{jm}\,\mathbf{M}_{\omega^{\prime}\beta m}(\rho_{j}%
)\Bigr),\medskip\\
\displaystyle\mathbf{H}_{j}(\mathbf{r})=e^{-i\omega t}\,e^{i\beta z}n^{0}%
\sum\limits_{m}e^{im\phi_{j}}\,\Bigl(a_{jm}\,\mathbf{M}_{\omega^{\prime}\beta
m}(\rho_{j})+b_{jm}\,\mathbf{N}_{\omega^{\prime}\beta m}(\rho_{j}%
)\Bigr),\quad\rho_{j}>R.
\end{array}
\label{EX_J}%
\end{equation}
Here another kind of the vector cylinder harmonics is introduced%

\begin{equation}
\mathbf{N}_{\omega^{\prime}\beta m}(\rho_{j})=\mathbf{e}_{r}\,\frac{i\beta
}{\varkappa^{\prime}}\,H_{m}^{\prime}(\varkappa^{\prime}\rho_{j}%
)-\mathbf{e}_{\phi}\,\frac{m\beta}{\varkappa^{\prime2}\rho_{j}}\,H_{m}%
(\varkappa^{\prime}\rho_{j})+\mathbf{e}_{z}\,H_{m}(\varkappa^{\prime}\rho
_{j}), \label{N}%
\end{equation}

\begin{equation}
\mathbf{M}_{\omega^{\prime}\beta m}(\rho_{j})=\mathbf{e}_{r}\,\frac
{m\omega^{\prime}}{\varkappa^{\prime2}\rho_{j}}\,H_{m}(\varkappa^{\prime}%
\rho_{j})+\mathbf{e}_{\phi}\,\frac{i\omega^{\prime}}{\varkappa^{\prime}%
}\,H_{m}^{\prime}(\varkappa^{\prime}\rho_{j}),\label{M}%
\end{equation}
where $H_{m}(\varkappa^{\prime}\rho_{j})$ is the Hankel function of the first
kind, $\omega^{\prime}=n^{0}\omega$, $\varkappa^{\prime}=\sqrt{\omega
^{\prime2}-\beta^{2}}$, and $n^{0}$ is the refractive index of the
environment. The functions $\mathbf{N}$ and $\mathbf{M}$ are orthogonal. Thus,
the contribution of the $j$-th waveguide into the guided mode field outside
the array is defined both by the propagation constant $\beta$ and by the set
of the partial amplitudes $a_{jm}$, $b_{jm}$. Thus, the total field outside
the array is%

\begin{equation}
\displaystyle\mathbf{E}(\mathbf{r})=\sum\limits_{j=1}^{N}\mathbf{E}%
_{j}(\mathbf{r}),\qquad\displaystyle\mathbf{H}(\mathbf{r})=\sum\limits_{j=1}%
^{N}\mathbf{H}_{j}(\mathbf{r}).\label{Sum11}%
\end{equation}
Note that, for $\beta=0$, Eq. (\ref{AppS_EHint}) and Eq. (\ref{EX_J})
transform into the corresponding expressions in Ref. \cite{VanDeHulst},
however different notations are used there. Below, the factor $e^{-i\omega
t}\,e^{i\beta z}$ is omitted, for brevity.

To derive the set of equations which determines the partial amplitudes
$a_{jm}$, $b_{jm}$, $c_{jm}$, $d_{jm},$ one should take into account that the
fields $\widetilde{\mathbf{E}}_{j}(\mathbf{R}_{j})$, $\widetilde{\mathbf{H}%
}_{j}(\mathbf{R}_{j}),$ described by Eq. (\ref{AppS_EHint}) and the field
$\mathbf{E}(\mathbf{R}_{j})$, $\mathbf{H}(\mathbf{R}_{j}),$ described by Eq.
(\ref{Sum11}), are connected by the boundary conditions on the surface of each
waveguide of the array; here $\mathbf{R}_{j}$ be the radius-vector of a point
on the surface of the $j$-th waveguide. These fields are connected by the six
boundary conditions. However, only four of them are independent. It is
convenient to choose the four ones which connect the $\phi$- and the
$z$-components of the field. Thus, if the permeability of the waveguide
material and the environment is unity, one has%

\begin{equation}%
\begin{array}
[c]{c}%
\displaystyle\left(  \mathbf{E}(\mathbf{R}_{j})\right)  _{\phi}=\left(
\mathbf{\tilde{E}}_{j}(\mathbf{R}_{j})\right)  _{\phi},\qquad\left(
\mathbf{H}(\mathbf{R}_{j})\right)  _{\phi}=\left(  \mathbf{\tilde{H}}%
_{j}(\mathbf{R}_{j})\right)  _{\phi},\\
\displaystyle\left(  \mathbf{E}(\mathbf{R}_{j})\right)  _{z}=\left(
\mathbf{\tilde{E}}_{j}(\mathbf{R}_{j})\right)  _{z},\qquad\left(
\mathbf{H}(\mathbf{R}_{j})\right)  _{z}=\left(  \mathbf{\tilde{H}}%
_{j}(\mathbf{R}_{j})\right)  _{z},
\end{array}%
\begin{array}
[c]{c}%
\\
\qquad j=1,2,...,N.
\end{array}
\label{AppS_BoundCond}%
\end{equation}
Based on Eq. (\ref{AppS_BoundCond}), one can obtain the uniform system of the
linear equations with respect to the variables $a_{jm}$, $b_{jm},$ $c_{jm}$,
$d_{jm}$. As is shown in Appendix, the system of equations is decoupled. The
amplitudes $a_{jm}$, $b_{jm}$ obey the system of equations
\begin{equation}
\hat{S}_{jm}^{-1}\,\left(
\begin{matrix}
a_{jm}\\
b_{jm}%
\end{matrix}
\right)  -\sum\limits_{l\neq j}^{N}\,\sum\limits_{n=-\infty}^{\infty}%
\,U_{nm}^{lj}\left(
\begin{matrix}
a_{ln}\\
b_{ln}%
\end{matrix}
\right)  =0.\label{MainSyst*}%
\end{equation}
while the amplitudes $c_{jm}$, $d_{jm}$ are expressed in terms of the
amplitudes $a_{jm},b_{jm}$ (see (\ref{AppS_cd=Tab}) of Appendix). The explicit
expressions for the $2\times2$ matrixes $\hat{S}_{jm}^{-1}$ and $U_{nm}^{lj}$
are presented in Appendix. The uniform linear system of equations
(\ref{MainSyst*}) has a nontrivial solution if its principal determinant,
dependent on $\hat{S}_{jm}^{-1}\left(  \omega,\beta\right)  $ and $U_{nm}%
^{lj}\left(  \omega,\beta\right)  ,$ vanishes. This condition determines the
dispersion curve $\beta\left(  \omega\right)  $ implicitly.

The physical interpretation of Eq. (\ref{MainSyst*}) is the following. If the
interaction between the waveguides $U_{nm}^{lj}$ is neglected, each waveguide
of the array behaves as an isolated. In this case a nontrivial solution for
Eq. (\ref{MainSyst*}) exists if $\det\,\hat{S}_{jm}^{-1}=0$ at least for one
pair of the parameters $(j,m)$. This condition determines the set of the
propagation constants $\beta_{jm}^{\left(  0\right)  }$ as a function of the
frequency $\omega$, which gives rise to a guided mode characterized by orbital
number $m$ and connected with the isolated waveguide $j$ \cite{Marcuse}. If
all the isolated waveguides are identical, each one has the same propagation
constant $\beta_{jm}^{\left(  0\right)  }=\beta_{m}^{\left(  0\right)  }$. In
this case, the guided modes with the frequency $\omega$ and the orbital number
$m$ are $N$-fold degenerated guided modes. Taking into account the interaction
$U_{nm}^{lj}$ results in the formation of the $N$ hybridized modes, each one
characterizing with certain propagation constant $\beta_{km}$, $1\leq k\leq
N$. However, the $N$-fold degeneration in the propagation constant remains,
since all the hybridized modes possess the same frequency $\omega$. The values
of the propagation constants $\beta_{km}$ belong to a certain band centered
around the value $\beta_{m}^{\left(  0\right)  }$. If the array is
equidistant, each of these hybridized modes is characterized by one of $N$
quasi-wave vectors $k_{x}$ which belong to the Brillouin band $\left(
-\pi/a,\pi/a\right)  $. Thus, for a given frequency, the guided modes possess
a certain dependence $\beta\left(  k_{x}\right)  $ which determines the
isofrequency curve. Naturally, if the refractive indices of the waveguides
differ or the array is not equidistant, this feature of the guided modes does
not hold for.

However, if the distance between the waveguides or their refractive index
fluctuates weakly, one may still have the $N$-fold degenerated guided mode
degenerated in the propagation constant. In the next section, we consider the
case of the equidistant array for which $\left(  n_{j}-n_{j+1}\right)
=\mathrm{const}\ll n_{j}$. In this case the propagation constants for the
isolated waveguides obey the relation $\beta_{j}^{\left(  0\right)  }%
=\beta_{0}^{(0)}+\alpha\cdot j$, $\alpha\ll\beta_{0}$.

\section{The nearest neighbor and the zero-harmonic approximation}

Even if the number of the waveguides in the array is finite, the system of
equations (\ref{MainSyst*}) is infinite since the number of different
harmonics $m$ remains infinite. Below we consider the simplest approximation
to these equations, namely, the harmonics with $m=0$ alone contribute to the
guided modes. Then, in the nearest neighbor approximation Eq. (\ref{MainSyst*}%
) takes the explicit form%
\begin{align}
\frac{a_{j}}{\overline{a}_{j}\left(  \beta\right)  }-\Bigl(U_{j}^{j+1}%
a_{j+1}+U_{j}^{j-1}a_{j-1}\Bigr) &  =0,\label{ms1}\\
\frac{b_{j}}{\overline{b}_{j}\left(  \beta\right)  }-\Bigl(U_{j}^{j+1}%
b_{j+1}+U_{j}^{j-1}b_{j-1}\Bigr) &  =0,\nonumber
\end{align}
where $a_{j}=a_{j0}$, $b_{j}=b_{j0}$, $U_{j}^{j\pm1}=U_{0,0}^{j\pm1,\,j}$,%
\begin{align}
\frac{1}{\overline{a}_{j}(\beta)} &  =\frac{\varepsilon^{0}\varkappa_{j}%
J_{0}\left(  \varkappa_{j}~R\right)  H_{0}^{\prime}\left(  \varkappa^{\prime
}R\right)  -\varepsilon_{j}\varkappa^{\prime}J_{0}^{\prime}\left(
\varkappa_{j}~R\right)  H_{0}\left(  \varkappa^{\prime}R\right)  }%
{\varepsilon_{j}\varkappa^{\prime}~J_{0}^{\prime}\left(  \varkappa
_{j}~R\right)  J_{0}\left(  \varkappa_{j}R\right)  -\varepsilon^{0}%
\varkappa_{j}J_{0}\left(  \varkappa_{j}~R\right)  J_{0}^{\prime}\left(
\varkappa_{j}R\right)  },\label{8}\\
\frac{1}{\overline{b}_{j}\left(  \beta\right)  } &  =\frac{\varkappa_{j}%
~J_{0}\left(  \varkappa_{j}~R\right)  H_{0}^{\prime}\left(  \varkappa^{\prime
}R\right)  -\varkappa^{\prime}J_{0}^{\prime}\left(  \varkappa_{j}~R\right)
H_{0}\left(  \varkappa^{\prime}R\right)  }{\varkappa^{\prime}~J_{0}^{\prime
}\left(  \varkappa_{j}~R\right)  J_{0}\left(  \varkappa_{j}R\right)
-\varkappa_{j}J_{0}\left(  \varkappa_{j}~R\right)  J_{0}^{\prime}\left(
\varkappa_{j}R\right)  },\nonumber
\end{align}
and $\varepsilon^{0}=\left(  n^{0}\right)  ^{2}$, $\varepsilon_{j}=n_{j}^{2}$.
The poles of $\overline{a}_{j}(\beta)$ and $\overline{b}_{j}\left(
\beta\right)  $ determine the guided modes for isolated waveguides.

First of all, let us note that for $m=0$ the system of equations (\ref{ms1})
decouples into two ones which describe the $a$- and the $b$- modes. Let us
consider, for example, the features of the modes ($b_{j}\equiv0$). In this
case, we deal with the uniform linear system of equations with respect to the
$N$ variables $a_{j}$. Then, a nontrivial solution of Eq. (\ref{ms1}) exists
if%
\begin{equation}
\det\left\vert \frac{1}{\overline{a}_{j}\left(  \beta\right)  }\delta
_{ij}+U_{i}^{j}\left(  \delta_{i,\,j-1}+\delta_{i,\,j+1}\right)  \right\vert
=0.\label{9-1}%
\end{equation}

Let the number of the waveguides $N$ be finite and $N\gg1$. Then, Eq.
(\ref{9-1}) determines several different values $\beta_{k}=\beta_{k}\left(
\omega\right)  $. It is easy to see that, if $\left(  n_{j}-n_{j+1}\right)
=\mathrm{const}\ll n_{j}$, there are $N$ solutions of Eq. (\ref{9-1}). Note
that each propagation constant $\beta_{k}$ determines the guided mode with the
same frequency $\omega$. Let $a_{j}\left(  \beta_{k}\right)  $ be the
normalized solution of Eq. (\ref{ms1}) and $\sum\nolimits_{j}\left\vert
a_{j}\left(  \beta_{k}\right)  \right\vert ^{2}=1$. Then, the monochromatic
guided mode, in the general case, is a linear superposition of the modes with
the different $\beta_{k}$:
\begin{align*}
\mathbf{E}\left(  t,\mathbf{r}\right)   &  =e^{-i\omega t}\sum_{k}%
C_{k}\,e^{i\beta_{k}z}\sum_{j=1}^{N}a_{j}\left(  \beta_{k}\right)
\mathbf{N}_{\omega^{\prime}\beta_{k}0}(\rho_{j}),\\
\mathbf{H}\left(  t,\mathbf{r}\right)   &  =e^{-i\omega t}n^{0}\sum_{k}%
C_{k}\,e^{i\beta_{k}z}\sum_{j=1}^{N}a_{j}\left(  \beta_{k}\right)
\mathbf{M}_{\omega^{\prime}\beta_{k}0}(\rho_{j}).
\end{align*}
The factors $C_{k}$ determine the linear superposition.

Let us introduce the modal amplitude%

\begin{equation}
a_{j}\left(  z\right)  =\sum_{k}C_{k}\,e^{i\beta_{k}z}\,a_{j}\left(  \beta
_{k}\right)  . \label{iii}%
\end{equation}
Since the functions $\mathbf{N}_{\omega^{\prime}\beta_{i}0}(\rho_{j})$,
$\mathbf{M}_{\omega^{\prime}\beta_{i}0}(\rho_{j})$ vanish rapidly as $\rho
_{j}$ increases, the field near the $j$-th waveguide is mainly determined by
the partial amplitudes $a_{j}\left(  \beta_{k}\right)  $. For this reason, the
modal amplitude $a_{j}\left(  z\right)  $ represents the behavior of the
guided modes properly. The coefficients $C_{k}$ are obtained from the boundary
condition at $z=0$:%

\begin{equation}
a_{j}(0)=\sum\limits_{k}\,C_{k}\,a_{j}(\beta_{k}),\label{a4}%
\end{equation}
$a_{j}(0)$ being given. The number of the different coefficients $C_{k}$
coincides with the total number of the waveguides in the array $N,$ the number
of equations in (\ref{a4}). In what follows, we assume the Gaussian form for
the modal amplitude behavior at $z=0$, i.e.%

\begin{equation}
a_{j}(0)=e^{-\frac{(j-j_{0})^{2}}{\sigma^{2}}+ik_{0}aj}. \label{a5}%
\end{equation}
This means that the external source approximately illuminates the ends of the
waveguides with the numbers $j_{0}-\sigma<j$ $<j_{0}+\sigma$ and the phase
difference between the amplitudes taken at the ends of the nearest waveguides
is $k_{0}a$.

Thus, to find the guided mode for the array under consideration, one should
perform the sequence of operations, namely: using Eq. (\ref{9-1}) calculate
numerically the set of propagating constants $\beta_{k}$; using Eq.
(\ref{ms1}), calculate the amplitudes $a_{j}(\beta_{k})$; specify the
distribution of the mode amplitude in the cross-section $z=0$, determined by
the parameters $\sigma$, $j_{0}$, $k_{0}$ (see Eq. (\ref{a5})); using Eq.
(\ref{a4}), calculate the coefficients $C_{k}$. This completely determines the
function $a_{j}\left(  z\right)  $.

\section{Justification and derivation of the phenomenological model}

It is demonstrated in the previous section how to calculate the modal
amplitude $a_{j}\left(  z\right)  $ which describes the guided modes. Let us
show that, under sufficiently general assumption, these amplitudes obey
phenomenological equation (\ref{intro}). Suppose that the refractive index is
$n_{j}=n_{0}+j\cdot\delta n$, $\delta n\ll n_{0}$. Then, the solutions of the
equation $\frac{1}{\overline{a}_{j}\left(  \beta\right)  }=0$ for different
$j$ which determine the propagating constants $\beta_{j}^{\left(  0\right)  }$
may be represented in the following form:
\begin{equation}
\beta_{j}^{\left(  0\right)  }=\beta_{0}^{\left(  0\right)  }+\alpha\cdot
j,\qquad\alpha\ll\beta_{0}^{(0)}.\label{assumption}%
\end{equation}
Here $\alpha$ is the parameter which determines the ramp in $\beta
_{j}^{\left(  0\right)  }$. If the coupling $U_{j}^{j\pm1}$ is weak enough,
the set of different solutions of Eq. (\ref{9-1}) $\beta_{k}$ obeys the
condition $\left\vert \left(  \beta_{k}-\beta_{j}^{\left(  0\right)  }\right)
/\beta_{j}^{\left(  0\right)  }\right\vert \ll1$. Then, since%
\begin{equation}
\frac{1}{\overline{a}_{j}\left(  \beta_{j}^{\left(  0\right)  }\right)
}=0,\label{9-2-1}%
\end{equation}
one has%
\begin{equation}
\frac{1}{\overline{a}_{j}\left(  \beta_{k}\right)  }\approx\left.
\frac{\partial}{\partial\beta}\frac{1}{\overline{a}_{j}\left(  \beta\right)
}\right\vert _{\beta=\beta_{j}^{\left(  0\right)  }}\cdot\left(  \beta
_{k}-\beta_{j}^{\left(  0\right)  }\right)  .\label{9-2}%
\end{equation}
Within the same accuracy one can assume that the parameter
\begin{equation}
\gamma_{j}\left(  \beta_{k}\right)  =U_{j}^{j\pm1}\left(  \beta_{k}\right)
/\left(  \left.  \frac{\partial}{\partial\beta}\frac{1}{\overline{a}%
_{j}\left(  \beta\right)  }\right\vert _{\beta=\beta_{j}^{\left(  0\right)  }%
}\right)  =\gamma\label{9}%
\end{equation}
depends weakly both on the number $j$ and on the value of the parameter
$\beta_{k}$. (The correctness of (\ref{9-2}) and (\ref{9}) can easily be
verified for any specific physical parameters describing the waveguide array).
Then, Eq. (\ref{ms1}) goes over to the following one%

\begin{equation}
\left(  \beta-\beta_{j}^{\left(  0\right)  }\right)  a_{j}-\gamma
\Bigl(a_{j-1}+a_{j+1}\Bigr)=0.\label{10}%
\end{equation}
This is a uniform system of linear equations with respect to the
variable $a_{j}$. Let $\widetilde{\beta}_{k}$ be set of the
different values resulting in a nontrivial solution of Eq.
(\ref{10}). As one expects, the number of $\widetilde{\beta}_{k}$
is $N$, the number of the waveguides in the array.
They are distributed within the interval $\left\vert \widetilde{\beta}%
_{k}-\beta_{0}^{\left(  0\right)  }\right\vert \leq\gamma$. Thus, the
approximate equation Eq. (\ref{10}) is valid if
\begin{equation}
\alpha\ll\gamma\ll\beta_{0}^{\left(  0\right)  }.\label{valid}%
\end{equation}
Let us introduce the modal amplitude%

\[
\widetilde{a}_{j}\left(  z\right)  =\sum_{k}\widetilde{C}_{k}\,e^{i\widetilde
{\beta}_{k}z}\,a_{j}(\widetilde{\beta}_{k}).
\]
Like modal amplitude (\ref{iii}), the amplitude $\widetilde{a}_{j}\left(
z\right)  $ represents the monochromatic guided mode properly. It is easy to
verify that these amplitudes satisfy the equation%
\begin{equation}
\left(  i\frac{\partial}{\partial z}+\beta_{j}^{\left(  0\right)  }\right)
\widetilde{a}_{j}\left(  z\right)  +\gamma\Bigl(\widetilde{a}_{j+1}\left(
z\right)  +\widetilde{a}_{j-1}\left(  z\right)  \Bigr)=0.\label{master}%
\end{equation}
Under assumption (\ref{assumption}) one obtains
\begin{equation}
\left(  i\frac{\partial}{\partial z}+\beta_{0}^{\left(  0\right)  }%
+\alpha\cdot j\right)  \widetilde{a}_{j}\left(  z\right)  +\gamma
\Bigl(\widetilde{a}_{j+1}\left(  z\right)  +\widetilde{a}_{j-1}\left(
z\right)  \Bigr)=0.\label{master1}%
\end{equation}
One can remove the constant $\beta_{0}^{\left(  0\right)  }~$from the last
equation by means of the phase calibration of the modal amplitudes
$\widetilde{a}_{j}\left(  z\right)  $. Then, the equation obtained coincides
with Eq. (\ref{intro}).

The solution of Eq. (\ref{master1}) can analytically be obtained
in the case 
$\sigma\gg1$ (see boundary condition
(\ref{a5})).
%
%
%
The solution of Eq. (\ref{master1}) takes on the form
\begin{equation}
\widetilde{a}_{j}(z)\approx e^{i\alpha(j-j_{0})z+i\phi(z)}\,e^{-\frac
{(j-j_{0}-\delta j(z))^{2}}{\sigma^{2}}+ik_{0}\,(j-j_{0}-\delta j(z))}%
,\label{b4}%
\end{equation}
where%

\begin{equation}
\delta j(z)=\frac{2\gamma}{\alpha}\Bigl(\cos\,k_{0}-\cos(k_{0}+\alpha
\,z)\Bigr), \label{b5}%
\end{equation}
and%

\begin{equation}
\phi(z)=-i\frac{2\gamma}{\alpha}\,\Bigl[\Bigl(\sin\,k_{0}-\sin(k_{0}%
+\alpha\,z)\Bigr)-k_{0}\,\Bigl(\cos\,k_{0}-\cos(k_{0}+\alpha\,z)\Bigr)\Bigr].
\label{b6}%
\end{equation}

The behavior of the factor $e^{-\frac{(j-j_{0}-\delta j(z))^{2}}{\sigma^{2}}}$
is the most interesting feature of the solution obtained. For $z=0$, the
function $\delta j(z)$ vanishes and the amplitude $\widetilde{a}_{j}(0)$ has a
noticeable value for the waveguides with the numbers $j_{0}-\sigma\leq j\leq
j_{0}+\sigma$. However, for $z>0$, $\delta j(z)\neq0$ and the factor
$e^{-\frac{(j-j_{0}-\delta j(z))^{2}}{\sigma^{2}}}$ describes the shift of the
numbers of the waveguides where $\widetilde{a}_{j}(z)$ possesses a noticeable
value. The oscillating dependence of the factor $\delta j(z)$ manifests the
Bloch oscillations. The function $\widetilde{a}_{j}(z)$ obtained describes the
distribution of the intensity in the space $j,\,z$. This distribution
possesses the maximal value along the sinusoidal trajectory with the period
$2\pi/\alpha$, while the amplitude excursion of the excitation is
$2\gamma/\alpha$.

Below we apply the results obtained to the experiments described in Refs.
\cite{87, Pertsch2}

\section{Application to the experiment}

To describe the guided modes for the array under consideration, on one hand,
one must turn to Eq. (\ref{ms1}). To obtain the solution of this equation, one
should first numerically solve Eq. (\ref{9-1}) to find the set $\beta_{k}$.
Given the parameters $k_{0}$ and $\sigma$, one obtains the amplitude
$a_{j}\left(  z\right)  $ (see Eq. (\ref{iii})). On the other hand, the
optical properties of the waveguide array under consideration can be described
by the amplitude $\widetilde{a}_{j}(z)$ (see Eq. (\ref{b4})). Thus the guided
modes may be described either by the function $a_{j}\left(  z\right)  $ or by
the function $\widetilde{a}_{j}(z)$. Let us apply the results obtained in the
two previous sections to the experiment described in Refs. \cite{87,
Pertsch2}. In these papers it has been revealed that the guided modes can
propagate 
as Bloch oscillations.

Let us show that the theory proposed agrees with the experiments and numerical
simulations in Refs. \cite{87, Pertsch2}. In these papers, the wavelength of
the laser source is $\lambda=633\,\mathrm{nm}$. The experiments are performed
for the homogeneous array of the waveguides in an inorganic-organic polymer
(the refractive index $n_{co}=1.554$) on the glass wafers (the refractive
index $n_{sub}=1.457$) with polymer cladding (the refractive index
$n_{cl}=1.550$). Each waveguide has a cross-section of $3.5\times
3.5\,\mathrm{\mu m}^{2}$. The uniform separation of the adjacent waveguides is
$8.5\,\mathrm{\mu m}$, the length of waveguides in the array is
$L=4.5\,\mathrm{cm}$. The uniform array is laterally detuned by taking an
advantage of the thermooptical effect in the polymer (the thermooptical
coefficient $n_{th}=10^{-4}\,\mathrm{K}^{-1}$). By the simultaneous heating
and cooling of the opposite sides, a lateral temperature gradient is
established, leading to a linear variation of the propagation constants of the
individual waveguides. The number of the waveguides in the array $N=75$. The
maximal total temperature drop is $\Delta T=25\,\mathrm{K}$. This drop results
in the maximal value $\alpha=250\,\mathrm{m}^{-1}$ within the experiment conditions.

We simulate the optical waveguide array studied in \cite{87, Pertsch2} by the
array of the evenly spaced identical cylinder waveguides as shown in Fig.
\ref{fig1}. The optical and geometrical parameters of the array we consider
are close to the parameters of the experiments. We assume that the waveguide
radius $R=1.975\,\mathrm{\mu m}$. This results in the waveguide cross-section
close to that in the experiment. The separation between the waveguides
$a=3R=5.925\,\mathrm{\mu m}$, what approximately corresponds with the
experiments. In the lack of the temperature gradient, the refractive index of
the waveguides is $n=1.554$. It is assumed that the refractive index of the
environment is $n^{0}\approx0.99\cdot n\approx1.538$, being certain average
between $n_{sub}$ and $n_{cl}$. If the total temperature gradient $\Delta T$
and the number of the waveguides $N$ are given, the refractive index
$n_{j}=n_{0}+n_{th}\cdot\frac{\Delta T}{N}\cdot j$, $-N/2\leq j\leq N/2$.
These data are enough to determine the parameters entering Eqs. (\ref{ms1}),
(\ref{master1}). To obtain the solutions of these equations, one should first
solve Eq. (\ref{9-1}) to find the set $\beta_{k}$. As the initial
approximation to solve Eq. (\ref{9-1}), one can take $\beta=\beta_{0}^{(0)}$
which is the solution of Eq. (\ref{9-2-1}) for the isolated waveguide. For the
values of the parameters given above, one obtains $\beta_{0}^{(0)}%
=1.535\times10^{7}\,\mathrm{m}^{-1}$, and using Eq. (\ref{9}) one obtains
$\gamma=198\,\mathrm{m}^{-1}$.

To compare the results obtained above with the numerical simulation and the
experimental findings performed in Refs. \cite{87, Pertsch2}, we assume that
$k_{0}=0$ in Eq. (\ref{a5}) and the temperature gradient between the nearest
waveguides $\delta T=\Delta T/N=5\cdot10^{-2}K$. Such temperature gradient
gives $\alpha=44~$m$^{-1},$ agreeing with Refs. \cite{87, Pertsch2}.



\begin{figure}[ptbh]
\includegraphics[width=0.47\textwidth]{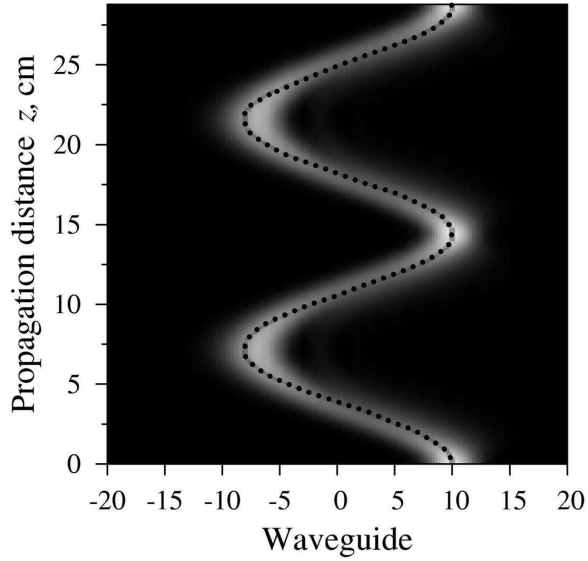}\caption{The
Bloch oscillations for the waveguide array. The brighter is the pixel in the
figure, the larger is the light intensity, described by the function
$a_{j}(z)$. The black dotted lines correspond to the position of maximal value
for the function $\widetilde{a}_{j}(z)$ (see Eq. (\ref{b4})).}%
\label{fig4}%
\end{figure}

To demonstrate the existence of the Bloch oscillations in the
waveguide array, one assumes that $\sigma=4$ in (\ref{a5}). Then,
the function $a_{j}\left( z\right)  $ describes these oscillations
as shown in Fig. \ref{fig4}. As mentioned above, both the
amplitude $a_{j}\left(  z\right)  $ ( solution of Eq. (\ref{iii}))
and the amplitude $\widetilde{a}_{j}\left(  z\right)  $ ( see Eq.
(\ref{master1})) can describe the effect. In order to compare
these results, the function $a_{j}\left(  z\right)  $ in Fig.
\ref{fig4} is represented completely by the areas of the different
brightness. However, the function $\widetilde{a}_{j}\left(
z\right)  $ is represented only partially by the bold dots where
it has the maximal value.

The result given in Fig. 
\ref{fig4} agrees with the numerical simulation in \cite{87,
Pertsch2}.

\begin{figure}[ptbh]
\includegraphics[width=0.47\textwidth]{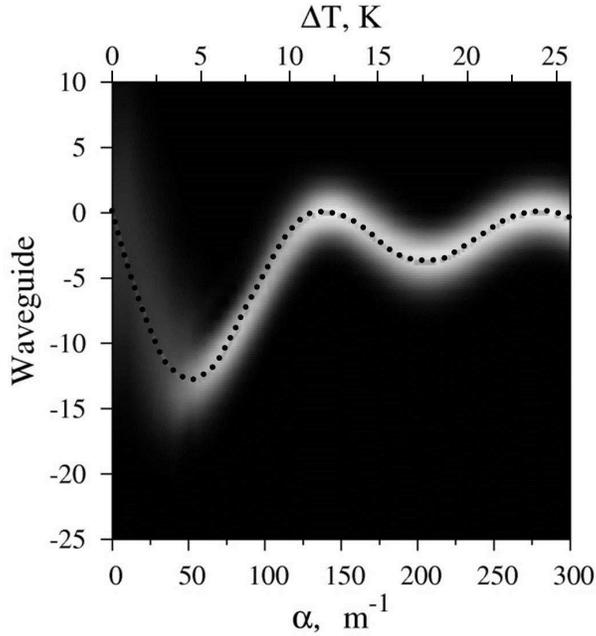}\caption{
The output intensity of the Bloch oscillations as function of the temperature
gradient $\Delta T$. The brighter is the pixel in the figure the larger is the
light intensity, described by function $a_{j}(L)$. The black dotted line
corresponds to the position of the maximal value for the function
$\widetilde{a}_{j}\left(  L\right)  $ (see Eq. (\ref{b4})).}%
\label{fig6}%
\end{figure}

To compare the results obtained with the experiment, one should investigate
the output intensity distribution as a function of the temperature gradient
$\Delta T$. Shown 
Fig. \ref{fig6} is the result of the simulation for
the Bloch oscillations, respectively for the parameters
corresponding to \cite{87, Pertsch2}. Since the parameter $\alpha$
is a single-valued function of $\Delta T$, 
Fig. \ref{fig6} manifests the output intensity distribution as a
function of the ramp in the propagation constant $\alpha$.


\section{Conclusion}

In this paper based on the macroscopic electrodynamics approach
and the multiple scattering formalism the system of equations is
derived describing the guided modes in the cylinder waveguide
array. This system contains the infinite number of equations,
being a formally exact one. The system of equations can be
truncated if one uses the nearest neighbor approximation and zero
harmonic approximation. In this case the system reduces to the
phenomenological description widely employed for recent decades.
Our approach allows us to calculate the unknown parameters which
determine the phenomenological equation. So far, these parameters
were extracted only as a result of comparison with the experiment.
It is found that the theory developed in this paper describes the
real experiments \cite{87, Pertsch2} in
which 
the Bloch oscillations are observed.
Surprisingly, not only qualitative but also reasonable quantitative agreement
is found.

Recently it has been communicated that the cylinder waveguide array considered
in this paper can be fabricated by means of the direct inscription of photonic
band-gap waveguides into bulk optical glass. Our theory may be applied to
these systems to attain the accurate description \cite{ICTON}.

\newpage

\subsection*{Acknowledgement.}

This work is supported by the Russian Federal Science and Innovation Program,
and the Russian Foundation for Basic Research. We are grateful to Yu. M. Kagan
for useful discussions and suggestions.


\section*{Appendix}

\setcounter{equation}{0} \renewcommand{\theequation}{A\arabic{equation}}

To use of the system of equations (\ref{AppS_BoundCond}), let us represent the
field $\mathbf{E}(\mathbf{R}_{j}),\mathbf{H}(\mathbf{R}_{j})$ in the form%
\begin{equation}%
\begin{array}
[c]{c}%
\displaystyle\mathbf{E}(\mathbf{R}_{j})=\mathbf{E}_{j}(\mathbf{R}_{j}%
)+\sum\limits_{l\neq j}^{N}\mathbf{E}_{l}(\mathbf{R}_{j}),\medskip\\
\displaystyle\mathbf{H}(\mathbf{R}_{j})=\mathbf{H}_{j}(\mathbf{R}_{j}%
)+\sum\limits_{l\neq j}^{N}\mathbf{H}_{l}(\mathbf{R}_{j}).
\end{array}
\label{Sum11-1}%
\end{equation}
Each of the fields $\mathbf{E}_{l}(\mathbf{R}_{j}),\mathbf{H}_{l}%
(\mathbf{R}_{j})$ is expressed in terms of the functions\textbf{\ }%
$\mathbf{N}\left(  \mathbf{R}_{j}-l\mathbf{a}\right)  $ and $\mathbf{M}\left(
\mathbf{R}_{j}-l\mathbf{a}\right)  $ (see Eqs. (\ref{EX_J})), i.e., these
functions are defined with respect to the \textit{different reference
systems}. Let us represent these functions for $l\neq j$ as linear expansions
in the functions $\tilde{\mathbf{N}}~\left(  \mathbf{R}_{j}-j\mathbf{a}%
\right)  ~$ and$~\tilde{\mathbf{M}}\left(  \mathbf{R}_{j}-j\mathbf{a}\right)
,$ i.e. the vector cylinder harmonics taken at\textit{\ the same reference
system}.

To do this let us employ the Graph theorem \cite{Graf}  given by the formula
\[
H_{n}(\rho_{l})\,e^{in\phi_{l}}=\sum\limits_{m=-\infty}^{+\infty}%
\,H_{n-m}(a|l-j|)\,J_{m}(\rho_{j})\,e^{im\phi_{j}},\quad\phi_{j}<\phi
_{l},\quad l\neq j.
\]
One can generalize this formula for arbitrary relation between the angles
$\phi_{j}$ and $\phi_{l}$. As a result, one obtains
\begin{equation}
H_{n}(\rho_{l})e^{in\phi_{l}}=\sum\limits_{m=-\infty}^{+\infty}\,H_{n-m}%
(a|l-j|)\,J_{m}(\rho_{j})\,e^{im\phi_{j}}\,\left[  \text{sign}(j-l)\right]
^{n-m},\quad l\neq j.\label{graf}%
\end{equation}
Using Eq. (\ref{graf}), one can show that
\begin{align}
e^{in\phi_{l}}\left(
\begin{array}
[c]{c}%
\mathbf{N}_{\omega^{\prime}\beta n}\left(  \rho_{l}\right)  \\
\mathbf{M}_{\omega^{\prime}\beta n}\left(  \rho_{l}\right)
\end{array}
\,\right)   &  =\sum\limits_{m=-\infty}^{+\infty}U_{nm}^{lj}\ e^{im\phi_{j}%
}\left(
\begin{array}
[c]{c}%
\tilde{\mathbf{N}}_{\omega^{\prime}\beta m}(\rho_{j})\\
\tilde{\mathbf{M}}_{\omega^{\prime}\beta m}(\rho_{j})
\end{array}
\right)  ,\medskip\label{grapf1}\\
\,U_{nm}^{lj}\, &  =H_{n-m}(\mathbf{\varkappa}^{\prime}a|l-j|)\left[
\text{sign}\left(  j-l\right)  \right]  ^{n-m},\qquad l\neq j.\nonumber
\end{align}

Taking these relations into account, one obtains%

\begin{equation}%
\begin{array}
[c]{l}%
\displaystyle\mathbf{E}_{l}(\mathbf{r})=\medskip\sum\limits_{m=-\infty
}^{\infty}e^{im\varphi_{j}}\left(  \sum\limits_{n=-\infty}^{\infty}%
\,\,U_{nm}^{lj}a_{ln}\right)  \tilde{\mathbf{N}}_{\omega^{\prime}\beta m}%
(\rho_{j})-\medskip\\
\displaystyle-\sum\limits_{m=-\infty}^{\infty}e^{im\phi_{j}}\left(
\sum\limits_{n=-\infty}^{\infty}\,\,U_{nm}^{lj}b_{ln}\right)  \tilde
{\mathbf{M}}_{\omega^{\prime}\beta m}(\rho_{j})\bigskip\\
\displaystyle\mathbf{H}_{l}(\mathbf{r})=\medskip n^{0}\sum\limits_{m=-\infty
}^{\infty}e^{im\phi_{j}}\left(  \sum\limits_{n=-\infty}^{\infty}%
\,\,U_{nm}^{lj}a_{ln}\right)  \tilde{\mathbf{M}}_{\omega^{\prime}\beta m}%
(\rho_{j})+\medskip\\
\displaystyle+n^{0}\sum\limits_{m=-\infty}^{\infty}e^{im\phi_{j}}\left(
\sum\limits_{n=-\infty}^{\infty}\,\,U_{nm}^{lj}b_{ln}\right)  \tilde
{\mathbf{N}}_{\omega^{\prime}\beta m}(\rho_{j}),\qquad\rho_{j}\geq R,\quad
l\neq j
\end{array}
\label{external1}%
\end{equation}

Let us replace $\mathbf{r}$ with $\mathbf{R}_{j}$ in Eqs. (\ref{external1}),
(\ref{AppS_EHint}), (\ref{EX_J}) and substitute them  into the system of
equations (\ref{AppS_BoundCond}). As a result, one obtains the uniform system
of linear equations for the partial amplitudes $a_{jm}$, $b_{jm}$, $c_{jm}$,
$d_{jm}$:%

\begin{equation}%
\begin{array}
[c]{c}%
\displaystyle a_{jm}\,\Bigl(N_{\omega^{\prime}\beta m\phi}\Bigr)-b_{jm}%
\,\Bigl(M_{\omega^{\prime}\beta m}\Bigr)_{\phi}+\left(  \sum\limits_{l\neq
j,~n}\,U_{nm}^{lj}\,a_{ln}\right)  \,\Bigl(\tilde{N}_{\omega^{\prime}\beta
m}\Bigr)_{\phi}-\medskip\\
\displaystyle-\left(  \sum\limits_{l\neq j,~n}\,U_{nm}^{lj}\,b_{ln}\right)
\,\Bigl(\tilde{M}_{\omega^{\prime}\beta m}\Bigr)_{\phi}=c_{jm}\,\Bigl(\tilde
{N}_{\omega_{j}\beta m}\Bigr)_{\phi}-d_{jm}\,\Bigl(\tilde{M}_{\omega_{j}\beta
m}\Bigr)_{\phi},
\end{array}
\label{Appx1}%
\end{equation}

\bigskip%

\begin{equation}%
\begin{array}
[c]{c}%
\displaystyle a_{jm}\,n^{0}\Bigl(M_{\omega^{\prime}\beta m}\Bigr)_{\phi
}+b_{jm}\,n^{0}\Bigl(N_{\omega^{\prime}\beta m}\Bigr)_{\phi}+\left(
\sum\limits_{l\neq j,~n}\,U_{nm}^{lj}a_{ln}\right)  \,n^{0}\Bigl(\tilde
{M}_{\omega^{\prime}\beta m}\Bigr)_{\phi}+\medskip\\
\displaystyle+\left(  \sum\limits_{l\neq j,~n}\,U_{nm}^{lj}\,b_{ln}\right)
\,n^{0}\Bigl(\tilde{N}_{\omega^{\prime}\beta m}\Bigr)_{\phi}=c_{jm}%
\,n_{j}\Bigl(\tilde{M}_{\omega_{j}\beta m}\Bigr)_{\phi}+d_{jm}\,n_{j}%
\Bigl(\tilde{N}_{\omega_{j}\beta m}\Bigr)_{\phi},
\end{array}
\label{Appx2}%
\end{equation}

\bigskip%

\begin{equation}%
\begin{array}
[c]{c}%
\displaystyle a_{jm}\,\Bigl(N_{\omega^{\prime}\beta m}\Bigr)_{z}+\left(
\sum\limits_{l\neq j,~n}\,U_{nm}^{lj}a_{ln}\right)  \,\Bigl(\tilde{N}%
_{\omega^{\prime}\beta m}\Bigr)_{z}=c_{jm}\,\Bigl(\tilde{N}_{\omega_{j}\beta
m}\Bigr)_{z},
\end{array}
\label{Appx3}%
\end{equation}

\bigskip%

\begin{equation}%
\begin{array}
[c]{c}%
\displaystyle b_{jm}\,n^{0}\Bigl(N_{\omega^{\prime}\beta m}\Bigr)_{z}+\left(
\sum\limits_{l\neq j,~n}\,U_{nm}^{lj}b_{ln}\right)  \,n^{0}\Bigl(\tilde
{N}_{\omega^{\prime}\beta m}\Bigr)_{z}=d_{jm}\,n_{j}\Bigl(\tilde{N}%
_{\omega_{j}\beta m}\Bigr)_{z}.
\end{array}
\label{Appx4}%
\end{equation}

Here, for brevity, the argument of the functions $N_{\omega^{\prime}\beta
m}(R),M_{\omega^{\prime}\beta m}(R),\tilde{N}_{\omega^{\prime}\beta
m}(R),\tilde{M}_{\omega^{\prime}\beta m}(R)$ is omitted. Here $R$ is the
radius of the rod, $n_{j}$ is the refractive index of the $j$-th rod, and
$n^{0}$ is the refractive index of the medium. In Eqs (\ref{Appx3}) and
(\ref{Appx4}), we have taken in account that $\Bigl(\tilde{M}_{\omega\beta
m}\Bigr)_{z}(\rho)=0$.

The system of equations (\ref{Appx1}) - (\ref{Appx4}) may be represented in
the form:%

\begin{equation}
\left(
\begin{matrix}
\hat{M}_{11} & \hat{M}_{12}\\
\hat{M}_{21} & \hat{M}_{22}%
\end{matrix}
\right)  \,\left(
\begin{matrix}
a_{jm}\\
b_{jm}\\
c_{jm}\\
d_{jm}%
\end{matrix}
\right)  =\left(
\begin{matrix}
\hat{N}_{1}\\
\hat{N}_{2}%
\end{matrix}
\right)  \,\left(
\begin{matrix}
\displaystyle\sum\limits_{l\neq j,~n}\,U_{nm}^{lj}a_{ln}\\
\displaystyle\sum\limits_{l\neq j,~n}\,U_{nm}^{lj}\,b_{ln}%
\end{matrix}
\right)  . \label{AppS_MatrEq1}%
\end{equation}

Here the matrix $\hat{M}_{11},\hat{M}_{12},\hat{M}_{21},\hat{M}_{22}\ $are
defind as follows%

\[
\hat{M}_{11}=\left(
\begin{matrix}
\Bigl(N_{\omega^{\prime}\beta m}\Bigr)_{\phi} & -\Bigl(M_{\omega^{\prime}\beta
m}\Bigr)_{\phi}\\
n^{0}\,\Bigl(M_{\omega^{\prime}\beta m}\Bigr)_{\phi} & n^{0}\,\Bigl(N_{\omega
^{\prime}\beta m}\Bigr)_{\phi}%
\end{matrix}
\right)  ,\qquad\hat{M}_{21}=\left(
\begin{matrix}
\Bigl(N_{\omega^{\prime}\beta m}\Bigr)_{z} & 0\\
0 & n^{0}\,\Bigl(N_{\omega^{\prime}\beta m}\Bigr)_{z}%
\end{matrix}
\right)  ,
\]%
\[
\hat{M}_{12}=\left(
\begin{matrix}
-\Bigl(\tilde{N}_{\omega_{j}\beta m}\Bigr)_{\phi} & \Bigl(\tilde{M}%
_{\omega_{j}\beta m}\Bigr)_{\phi}\\
-n_{j}\,\bigl(\tilde{M}_{\omega_{j}\beta m}\Bigr)_{\phi} & -n_{j}%
\,\Bigl(\tilde{N}_{\omega_{j}\beta m}\Bigr)_{\phi}%
\end{matrix}
\right)  ,\qquad\hat{M}_{22}=\left(
\begin{matrix}
-\Bigl(\tilde{N}_{\omega_{j}\beta m}\Bigr)_{z} & 0\\
0 & -n_{j}\,\Bigl(\tilde{N}_{\omega_{j}\beta m}\Bigr)_{z}%
\end{matrix}
\right)  ,
\]%
\[
\hat{N}_{1}=\left(
\begin{matrix}
-\Bigl(\tilde{N}_{\omega^{\prime}\beta m}\Bigr)_{\phi} & \Bigl(\tilde
{M}_{\omega^{\prime}\beta m}\Bigl)_{\phi}\\
-n\,\Bigl(\tilde{M}_{\omega^{\prime}\beta m}\Bigr)_{\phi} & -n\,\Bigl(\tilde
{N}_{\omega^{\prime}\beta m}\Bigr)_{\phi}%
\end{matrix}
\right)  ,\qquad\hat{N}_{2}=\left(
\begin{matrix}
-\Bigl(\tilde{N}_{\omega^{\prime}\beta m}\Bigr)_{z} & 0\\
0 & -n\,\Bigl(\tilde{N}_{\omega^{\prime}\beta m}\Bigr)_{z}%
\end{matrix}
\right)  ,
\]
Then, the system of equations (\ref{AppS_MatrEq1}) is transformed into the form%

\begin{gather}
\hat{S}_{jm}^{-1}\,\left(
\begin{matrix}
a_{jm}\\
b_{jm}%
\end{matrix}
\right)  -\sum\limits_{l\neq j}^{N}\,\sum\limits_{n=-\infty}^{+\infty}%
\,U_{lm}^{jn}\,\left(
\begin{matrix}
a_{ln}\\
b_{ln}%
\end{matrix}
\right)  =0,\label{MainSyst}\\
\left(
\begin{matrix}
c_{jm}\\
d_{jm}%
\end{matrix}
\right)  =\hat{T}_{jm}\left(
\begin{matrix}
a_{jm}\\
b_{jm}%
\end{matrix}
\right)  , \label{AppS_cd=Tab}%
\end{gather}
where%

\begin{equation}
\hat{S}_{jm}=\biggl(\hat{M}_{12}^{-1}\,\hat{M}_{11}-\hat{M}_{22}^{-1}\,\hat
{M}_{21}\biggr)^{-1}\biggl(\hat{M}_{12}^{-1}\,\hat{N}_{1}-\hat{M}_{22}%
^{-1}\,\hat{N}_{2}\biggr), \label{AppS_S}%
\end{equation}

\begin{equation}
\hat{T}_{mj}=-\biggl(\hat{N}_{1}^{-1}\hat{M}_{12}-\hat{N}_{2}^{-1}\hat{M}%
_{22}\biggr)^{-1}\biggl(\hat{N}_{1}^{-1}\hat{M}_{11}-\hat{N}_{2}^{-1}\hat
{M}_{21}\biggr). \label{AppS_T}%
\end{equation}

Thus, finding the solution of Eq. (\ref{Appx1})-(\ref{Appx4}) reduces to
solving Eq. (\ref{MainSyst}) with respect to $a_{jm}$, $b_{jm},$and next
calculation of the amplitudes $c_{jm},d_{jm}$ using Eq. (\ref{AppS_cd=Tab}).

\section*{}
\end{document}